\documentclass[conference]{IEEEtran}
\IEEEoverridecommandlockouts
\usepackage{cite}
\usepackage{amsmath,amssymb,amsfonts}
\usepackage{algorithmic}
\usepackage{graphicx}
\usepackage{textcomp}
\usepackage{xcolor}

\usepackage{comment}
\usepackage{todonotes}
\usepackage{nameref}

\usepackage{listings}
\usepackage{esvect}

\usepackage[hidelinks]{hyperref} %this must come first to hide colored boxes at links
\usepackage{orcidlink} %this must come afterwards
\let\oldtexttt\texttt
\renewcommand{\texttt}[1]{{\small\oldtexttt{#1}}}
\usepackage{graphicx}
\usepackage{eso-pic}
\def\BibTeX{{\rm B\kern-.05em{\sc i\kern-.025em b}\kern-.08em
    T\kern-.1667em\lower.7ex\hbox{E}\kern-.125emX}}
\begin{document}

\title{DAXiot: A Decentralized Authentication and Authorization Scheme for Dynamic IoT Networks\\
\thanks{This paper has been realized via funding from the IDunion project by German Ministry of Economic Affairs and Climate Action (BMWK), grant number 01MN21002K. The information and views set out in this publication are those of the authors and do not necessarily reflect the official opinion of the BMWK. Responsibility for the information and views expressed here lies entirely with the authors.}
}

\author{\IEEEauthorblockN{Artur Philipp \orcidlink{0000-0003-3785-3130}}
\IEEEauthorblockA{\textit{Service-centric Networking (SNET)} \\
\textit{Technische Universität Berlin}\\
Berlin, Germany \\
a.philipp@tu-berlin.de}
\and
\IEEEauthorblockN{Axel Küpper \orcidlink{0000-0002-4356-5613}}
\IEEEauthorblockA{\textit{Service-centric Networking (SNET)} \\
\textit{Technische Universität Berlin}\\
Berlin, Germany \\
axel.kuepper@tu-berlin.de}
}

\maketitle

\AddToShipoutPicture*{\small \sffamily\raisebox{1.2cm}{\hspace{1.8cm}\parbox{\textwidth}{This work has been submitted to the IEEE for possible publication. Copyright may be transferred without notice, after which this version may no longer be accessible.}}}

\begin{abstract}
Federated and decentralized networks supporting frequently changing system participants are a requirement for future Internet of Things (IoT) use cases. IoT devices and networks often lack adequate authentication and authorization mechanisms, resulting in insufficient privacy for entities in such systems. In this work we address both issues by designing a privacy preserving challenge-response style authentication and authorization scheme based on Decentralized Identifiers and Verifiable Credentials. Our solution allows a decentralized permission management of frequently changing network participants and supports authenticated encryption for data confidentiality. We demonstrate our solution in an MQTT 5.0 scenario and evaluate its security, privacy guarantees, and performance.
\end{abstract}


\section{Introduction}
\label{Introduction}
While the Internet of Things (IoT) market continues to grow steadily, achieving adequate security for IoT devices and networks remains a challenge that is yet not sufficiently addressed by device manufacturers and network operators. \cite{bib_schiller_landscape_iot_security}. 
The IoT landscape is a heterogeneous field and established IoT protocols such as MQTT \cite{bib_mqtt} do not include authentication and authorization schemes in their specifications but rely on other measures like X.509 certificates, which come with the burden of managing complex infrastructures. However, authentication and authorization of IoT devices is mandatory for successful IoT use cases. Proper access control is required to preserve privacy and security in such systems \cite{bib_maigne_centralized_distributed_between_IoT_Security}. Privacy and security become even more crucial when several independent organizations collaborate within multiple IoT networks to achieve common goals. Federated networks are a requirement for future IoT use cases and technological movements like "Industry 4.0" and "Made in China 2025" support this statement \cite{bib_industry40_state_of_art}. Use cases such as connected vehicles involve frequently changing entities from different organizations and domains. 
Therefore, authentication and authorization frameworks for such dynamic and cooperative networks need to be secure, to respect the privacy of each participating organization, and to support a decentralized governance model. \cite{bib_ravidas_AC_in_IoT_survey} 

We recognize Decentralized Identifiers (DID) and Verifiable Credentials (VC) as valuable tools in addressing the aforementioned challenges. DIDs are identifiers independent of any central authority while VCs are verifiable cryptographic claims suitable for authentication and authorization. DID Documents (DDOC) provide meta information about DIDs and are often stored on distributed systems such as blockchains.

In this work we present DAXiot: A decentralized authentication and authorization scheme for dynamic IoT networks. We leverage DIDs, DDOCs, VCs and their features by using them as identifiers and authorization tokens for IoT devices. Our solution can be applied to any IoT protocol supporting challenge-response based authentication. We demonstrate our implementation of DAXiot in a scenario with MQTT 5.0, a publish-subscribe protocol for constrained IoT devices: Devices publish messages to a broker that forwards it to subscribers. We use DAXiot for authentication and verification of publish / subscribe authorizations. Our contributions are:

\begin{itemize}
    \item{We design a decentralized challenge-response based authentication and authorization scheme for IoT using Decentralized Identifiers and Verifiable Credentials}
    \item{We allow changing system participants with low effort}
    \item{We preserve privacy with authenticated application layer encryption and selective disclosure of authorization data}
\end{itemize}

The remainder of this work is organized as follows. In section \ref{sec_related_work} we present related work. In section \ref{sec_background_and_preliminaries} we provide preliminaries for an overview of DAXiot in section \ref{sec_system_overview}. In section \ref{sec_systm_design_and_message_flow} we demonstrate our solution in a concrete scenario. Section \ref{sec_adding_removing_system_participants} describes how to add or remove system participants. We evaluate DAXiot in section \ref{sec_implementation_and_evaluation} and present our conclusion with future work in section \ref{sec_conclusion_and_future_work}.

\section{Related Work}
\label{sec_related_work}
Abubakar et al. propose an authentication and authorization scheme for MQTT devices in which DIDs of users, mobile and IoT devices need to be publicly registered on a blockchain \cite{bib_abubakar_blockchain_based_MQTT}. Users are authenticated by signing a challenge. Authentication is publicly traceable and takes up to 30 seconds. The authors do not detail how communication channels are secured but compare their solution with Transport Layer Security (TLS).

Dixit et al. define a decentralized IoT identity framework and evaluate it with two different blockchains \cite{bib_dixit_decentralized_it_framework}. In one case, VCs are stored publicly on the blockchain. Device URLs are always public. Mutual authentication and authorization between devices is accomplished through VC exchange. The authors do not detail how message confidentiality or authenticity is achieved, nor how their trust model works. 

Belchior et al. map the attributes of VCs to conventional access control policies \cite{bib_belchior_ssibac}. The systems processes 55,000 access requests per second, supports attribute and role-based access control, and provides user privacy by using Zero Knowledge Proofs (ZKP). The authors identify the access control engine verifying the VCs as single point of failure in their concept.

Fotiou et al. designed a capabilities-based access control system in which users authenticate to an authorization server and receive a privacy preserving ephemeral DID and a VC over TLS \cite{bib_fotiou_capabality_based_access}. The VC defines the authorized user actions for an IoT device. Devices are managed by owners and are configured with lists of trusted issuers. When a new issuer joins the system, all devices have to be updated.

All examined solutions require additional components like Authorization Server, Access Control Engines, or a blockchain and execute authorization out-of-band. Hence we focus on a solution that can be applied in-band and aim to avoid external additional components. Based on the insights from examined works, we define the following properties for our system:

\textbf{Security:} Device communication must ensure confidentiality, integrity, and authenticity. Device endpoints must not be public. Mutual device authentication must be possible. Device owner must only have control of their devices.

\textbf{Privacy:} Public registration of devices, device owners or users must not be required. Device interactions or VCs must not be visible to the public.

\textbf{Manageability:} The solution shall be applicable to multiple IoT protocols. Adding or removing system participants must come with low effort, no extra cost, and must not require all devices to be updated. The system shall be decentralized.

\textbf{Performance:} Authentication and authorization of devices shall happen instantly. 

In the following sections we introduce the tools we leveraged for achieving these goals with DAXiot.

\section{Preliminaries}
\label{sec_background_and_preliminaries}

\subsection{Decentralized Identifiers}
\label{sec_sub_dids}
Decentralized Identifiers (DID) are independent from any central authority, registry, or identity provider and bind $1..n$ public keys. \cite{bib_w3c_did_core}.
A DID subject is the entity controlling the DID and proves ownership by using a private key associated with one of the DID's public keys. DID Resolvers (R) are software that resolve DIDs to DID Documents (DDOC). DDOCs contain meta data such as a) public keys for verifying signatures created by the DID, b) public keys for performing DH key-agreements with the DID, or c) service endpoint URIs for providing (web) services. A DID has the form \texttt{did:<method><identifier>}. The DID \texttt{<method>} (DM) defines how a DDOC can be created, read, or modified, followed by the DID's unique \texttt{<identifier>}. More than 160 DMs are available \cite{bib_w3c_did_specification_registries}. One categorization of DMs is the storage location of DDOCs. \texttt{did:ethr} \cite{bib_did_ethr_method} stores DDOCs on the Ethereum Blockchain while \texttt{did:web} \cite{bib_did_web_method} defines the DDOC location at a \texttt{https://<domain>/.well-known/did.json} URL. \texttt{did:key} does not require storing a DDOC publicly since a \texttt{did:key} DID's \texttt{<identifier>} serves as an encoded representation of the DID's public key. Hence a \texttt{did:key} DID inherently carries a DDOC with the DID's public key \cite{bib_did_key_method}.
 
\subsection{Verifiable Credentials}
\label{sec_sub_vcs}
DID subjects become issuers when they use private keys of their DID for issuing VCs to other DIDs. VCs are digitally signed and cryptographically verifiable data containers with claims about a subject \cite{bib_w3c_vc_data_model_11}. When issuers issue VCs to subjects, subjects become holders and can present VCs to verifiers for proving authenticity of embedded claims. Verifiers resolve DDOCs of issuer DIDs by using a resolver, and verify the signature of VCs with public keys from issuer's DDOCs. Revocation Registries (RR) contain the status of VCs and allow issuers to revoke issued VCs. Verifiers check a VC status in RRs after verifying a VC's signature.

Different VC formats have distinct privacy features \cite{bib_vc_flavors_Article}. For example, a JWT VC does not allow to reveal only a subset of claims to a verifier (selective disclosure (SD)). Other formats like AnonCreds \cite{bib_anoncreds_specification} support SD but are computation-intensive, which does not fit to constrained devices. However, SD is a crucial feature, since it enables privacy preserving authentication that is required within collaborative IoT networks.

Selective Disclosure for JWTs (SD-JWT) are a new lightweight JWT based credential format, that supports SD by using salted hashes \cite{bib_sd_jwt}. We use SD-JWTs and define an \texttt{AuthorizationCredential} ($AC$) which contains authorization claims for subscribing and publishing to topics. Listing \ref{listing_ac_plain_claims} presents an example with issuer \texttt{did:web:issuer.com}, subject \texttt{did:key:z6...i8} and two claims: The first claim authorizes the subject to subscribe to topic \texttt{t1} and to publish to \texttt{t2} at \texttt{did:web:broker1.com}. The second claim authorizes publishing to topics \texttt{t3} and \texttt{t4} at \texttt{did:web:broker2.com}.

The issuer prepares for each claim a disclosure (an array with a salt, the claim key and claim value), hashes it (Listing \ref{listing_ac_disclosures}) and replaces the plain text claims by the hashes in  \texttt{\_sd}. The result is a SD-JWT (Listing \ref{listing_ac_hashed_claims}.) and is signed by the issuer. SD-JWT and disclosures are sent to the holder by the issuer. The holder presents the SD-JWT with the selected set of disclosures to be revealed to a verifier. The Verifier checks the SD-JWT's signature first, then recalculates and compares the disclosure hashes to verify the SD-JWT's authenticity.

\begin{minipage}{\linewidth}
\begin{lstlisting}[
    basicstyle=\footnotesize\ttfamily, %or \small or \footnotesize etc.
    caption={Authorization Credential with plain text claims.},
    captionpos=b,
    label=listing_ac_plain_claims
]
{   "iss":  "did:web:issuer.com",
    "sub":  "did:key:z6...i8",
   "type":  "AuthorizationCredential",
    "jti":  "AC_ID_123456789",
   `"did:web:broker1.com"`: 
            {"sub": ["t1"], "pub": ["t2"]},
   `"did:web:broker2.com"`: 
            {"pub": ["t3", "t4"]}  }

\end{lstlisting}
\end{minipage}

\begin{minipage}{\linewidth}
\begin{lstlisting}[
    basicstyle=\footnotesize\ttfamily, %or \small or \footnotesize etc.
    caption={Disclosures for hashed claims of SD-JWT.},
    captionpos=b,
    label=listing_ac_disclosures
]
//Disclosure Hash: `"1rQ...jPI"`     =     Hash of
[ "2GLC...RN9w",                         //salt &
  `"did:web:broker1.com"`,                 //key  &
  {"sub": ["t1"], "pub": ["t2"]} ]       //value

//Disclosure Hash: `"2zY...rSE"`     =     Hash of
[ "eluV...xA_A",                         //salt &
  `"did:web:broker2.com"`,                 //key  &
  {"pub": ["t3", "t4"]} ]                //value
\end{lstlisting}
\end{minipage}

\begin{minipage}{\linewidth}
\begin{lstlisting}[
    basicstyle=\footnotesize\ttfamily, %or \small or \footnotesize etc.
    caption={SD-JWT with hashed claims.},
    captionpos=b,
    label=listing_ac_hashed_claims
]
{   "iss":  "did:web:issuer.com",
    "sub":  "did:key:z6...i8",
   "type":  "AuthorizationCredential",
    "jti":  "AC_ID_123456789",
   ` "_sd"`:  [`"1rQ...jPI"`, `"2zY...rSE"`]  }
\end{lstlisting}
\end{minipage}

\section{System Overview}
\label{sec_system_overview}
DAXiot is a decentralized and privacy preserving challenge-response based authentication and authorization scheme. We use DIDs and DDOCs for mutual authentication and SD-JWTs for privacy-preserving authorization as presented in \ref{sec_sub_vcs}. 

We leverage Elliptic Curve Cryptography (ECC) \cite{bib_elliptic_curve_cryptography} and Diffie–Hellman (DH) key-agreements \cite{bib_Diffie_Hellman_Key_Exchange} for establishing shared secrets by using private and public keys of DIDs and provide message authenticity, confidentiality, and integrity. ECC keys are shorter than RSA keys, and provide equivalent security and performance \cite{bib_state_of_elliptic_curve_cryptography}. For Credential signatures we use Edwards-curve Digital Signature Algorithm (EdDSA) with ed25519 keys \cite{bib_curve25519} and derive from these x25519 keys, which can be used to perform Elliptic Curve DH (ECDH). This enables a sender and receiver to perform EdDSA and ECDH based on a single ed25519 key pair \cite{bib_conversion_ed25519_x25519}. 

We perform a sequence of ECDH-ES (ephemeral-static) and ECDH-1PU (One-Pass Unified Model) DH for establishing a shared session key. ECDH-ES (ephemeral-static) requires the sender to use an (anonymous) ephemeral key pair with the receiver's (known) static long-term key pair to compute a shared ECDH-ES key. ECDH-1PU adds one additional ECDH key-agreement between the static long-term keys of sender and receiver and enables two-way authenticated encryption without a sender being required to add a message signature \cite{bib_ecdh1pu}.

\begin{figure}[ht]
    \centerline{\includegraphics[width=0.5\textwidth]{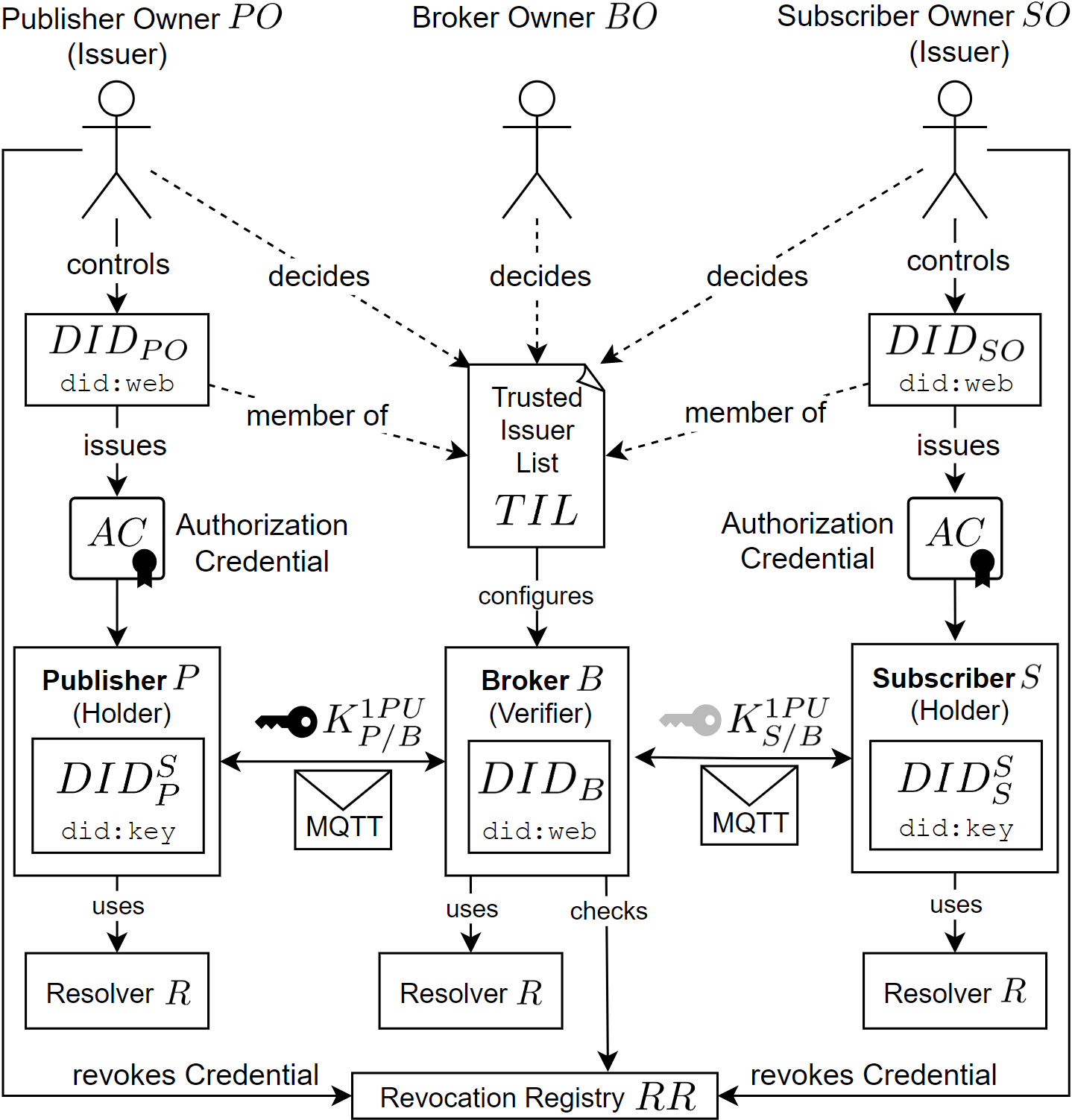}}
    \caption{DAXiot system overview with actors and components in MQTT 5.0.}
    \label{fig_system_overview}
\end{figure}

We encrypt a client's static long-term \texttt{did:key} with a computed ECDH-ES key, send it to the receiver and execute ECDH-1PU to compute the session key. We execute ECDH-1PU only once and not per message. As result, we loose Forward Secrecy but eliminate recurring computational effort. 
The following section provides an overview of actors and components 
in an MQTT 5.0 scenario with DAXiot (Fig. \ref{fig_system_overview}). 

\textbf{Broker} \pmb{$B$} is a MQTT 5.0 broker and is configured with a \texttt{did:web} $DID_B$ and a Trusted Issuer List $TIL$. $DID_B$ includes $B$'s service endpoint $SE_B$. $B$ receives \texttt{SUBSCRIBE} messages with an encrypted topic $T$ from subscriber $S$ and receives \texttt{PUBLISH} messages with an encrypted payload $PL$ and encrypted topic $T$ from publisher $P$. 
Authentication is conducted by $B$ when $P$ or $S$ connect via a \texttt{CONNECT} message and present $AC$s. $AC$s are processed by $B$ and authorization claims are stored by $B$. Authorization is checked by $B$ when $P$ publishes a message to $T$ or when $S$ subscribes to $T$. $B$ has the role of a verifier. $B$ stores every nonce used for encryption and verifies that it has not been used by the sender before.

\textbf{Broker Owner} \pmb{$BO$} configures $B$ with $TIL$ and $DID_B$.

\textbf{Publisher} \pmb{$P$} is a MQTT 5.0 client with a static long-term \texttt{did:key} $DID_P^S$ and $AC$. $P$ has the role of a holder. $P$ generates an ephemeral \texttt{did:key} $DID_P^E$ as \texttt{clientID} when connecting to $B$ and presents $AC$ to $B$.

\textbf{Publisher Owner} \pmb{$PO$} controls a \texttt{did:web} $DID_{PO}$ and configures $P$ with $AC$ and a static long-term \texttt{did:key} $DID_P^S$. $PO$ is an issuer and can mark issued $AC$s as \texttt{REVOKED} in {RR}.

\textbf{Subscriber} \pmb{$S$} is a MQTT 5.0 client with a static long-term \texttt{did:key} $DID_S^S$ and $AC$. $S$ is a holder and generates an ephemeral \texttt{did:key} $DID_S^E$ as \texttt{clientID} when connecting to $B$ and presents $AC$ to $B$. $S$ subscribes to topic $T$ at $B$ by sending a \texttt{SUBSCRIBE} message with encrypted $T$ to $B$.

\textbf{Subscriber Owner} \pmb{$SO$} controls a \texttt{did:web} $DID_{SO}$ and configures $S$ with $AC$ and a static long-term \texttt{did:key} $DID_S^S$. $SO$ is an issuer and can mark issued $AC$s as \texttt{REVOKED} in {RR}.

\textbf{Authorization Credential} \pmb{$AC$} is a signed SD-JWT issued by $PO$ to $P$ or $SO$ to $S$ and contains authorization claims. 

\textbf{Trusted Issuer List} \pmb{$TIL$} contains DIDs of trusted $PO$ and $SO$ defined by $BO$, $PO$ and $SO$. When $P$ or $S$ connects to $B$, $B$ verifies that the $AC$ of $P$ or $S$ is issued by a DID included in $TIL$. How $BO$, $PO$ and $SO$ consent on the inclusion or exclusion of DIDs in $TIL$ is out of scope in our work. 

$PO$ and $SO$ share a \textbf{Revocation Registry} \pmb{$RR$} for marking issued $AC$s as \texttt{REVOKED}. A verifier checks an $AC$'s status at $RR$ by using the id of the $AC$. $B$, $P$ and $S$ each have a \textbf{Resolver} \pmb{$R$}. These could also be included in $B$, $P$ and $S$.

\section{System Design and Message Flow}
\label{sec_systm_design_and_message_flow}

We demonstrate our system design and message flow by applying DAXiot to one concrete scenario with MQTT 5.0:

$B$ is configured with $TIL$. $DID_{PO}$ and $DID_{SO}$ are members of $TIL$. $P$ has an $AC$ from $PO$ that allows publishing to topic $T$ at $B$ and subscribing to $T_{other}$ at $B_{other}$. $S$ has an $AC$ from $SO$ that allows subscribing to $T$ at $B$. $P$ connects to $B$ and executes DAXiot with $B$. $P$ discloses only the authorization claim for publishing to $T$ at $B$. $S$ connects to $B$ and executes DAXiot with $B$. $S$ discloses the authorization claim for subscribing to $T$ at $B$. $S$ subscribes to $T$ at $B$. $P$ publishes a message to $T$ at $B$. $B$ receives the message, processes it and forwards it to $S$. Message payloads are encrypted with shared keys retrieved after performing DAXiot. 

\begin{figure}[ht]
    \centerline{\includegraphics[width=0.5\textwidth]{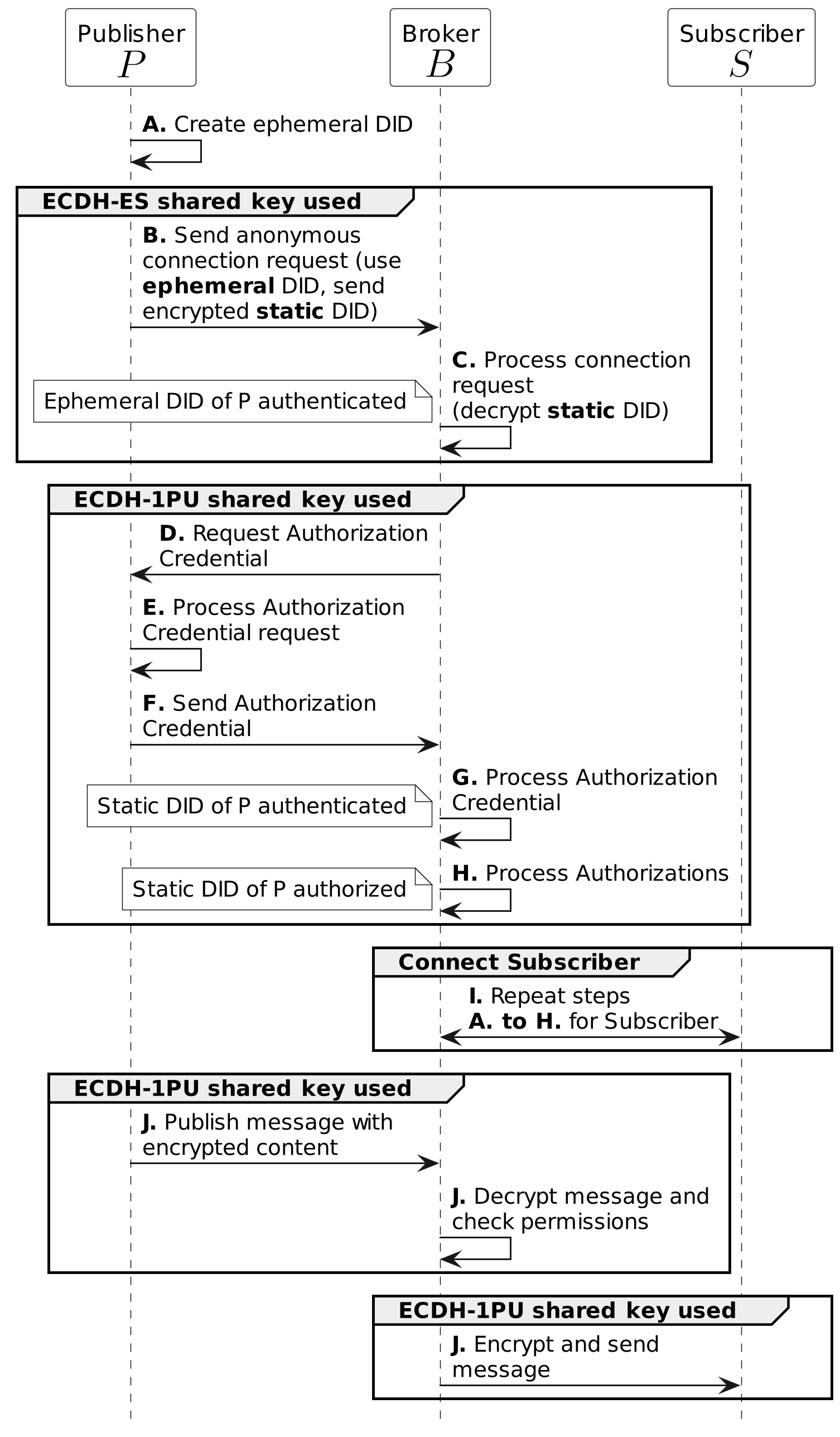}}
    \caption{DAXiot message flow between publisher, broker and subscriber.}
    \label{fig_ecdh1pu_flow}
\end{figure}

Fig. \ref{fig_ecdh1pu_flow} shows the complete flow of the scenario. Leading bold capital letters \textbf{A.} to \textbf{J.} in Fig. \ref{fig_ecdh1pu_flow} interactions correspond to the following sections, which detail the actions and cryptographic operations performed by $B$, $P$ and $S$. Message groups in Fig. \ref{fig_ecdh1pu_flow} indicate the key-agreement performed for data encryption.

\subsection{Publisher creates ephemeral DID}
$P$ generates an ephemeral ed25519 signature key pair consisting of the secret signature key $SK\_SIG_P^E$ and public signature key $PK\_SIG_P^E$ and derives from it the ephemeral x25519 secret key-agreement key $SK\_AGG_P^E$ and the ephemeral x25519 public key-agreement key $PK\_AGG_P^E$. $P$ uses multibase \cite{bib_multibase} to deterministically encode $PK\_SIG_P^E$ and creates a new ephemeral \texttt{did:key} $DID_P^E$ as follows: \texttt{did:key:\textless $mb(PK\_SIG_P^E)$\textgreater}. 

$P$ uses $DID_B$ from $AC$ and retrieves $B$'s DID Document $DDOC_B$ from $R$. $P$ extracts from $DDOC_B$ $B$'s service endpoint $SE_B$ as well as $B$'s multibase encoded x25519 public key-agreement key and decodes it ($PK\_AGG_B$). 
At this point, $P$ prepared all required data for sending an anonymous connection request to $B$.

\subsection{Publisher sends anonymous connection request}
$P$ uses its ephemeral secret key-agreement key $SK\_AGG_P^E$ and $B$'s public key-agreement key $PK\_AGG_B$ to execute ECDH-ES. The result is a shared key $K_{P/B}^{ES}$. $P$ generates a nonce $N$ and encrypts its static long-term ${DID_P^S}$ with $K_{P/B}^{ES}$ and $N$. The result is $enc({DID_P^S})$. $P$ prepares a \texttt{CONNECT} message $M_{con}$, sets ${DID_P^E}$ as \texttt{clientID}, sets $enc({DID_P^S})$ as \texttt{Authentication Data}, sets \texttt{DAXiot} as \texttt{Authentication Method}, and sends $M_{con}$ to $SE_B$. At this point, $P$ connected to $B$ using the (to $B$ yet unknown) ephemeral ${DID_P^E}$ as \texttt{clientID} and shared its static long-term $DID_P^S$. A man-in-the-middle (MitM) is not able to decrypt $DID_P^S$ from $M_{con}$.

\subsection{Broker processes connection request}
$B$ receives $M_{con}$ from $P$ and only proceeds when \texttt{Authentication Method} equals \texttt{DAXiot}. $B$ reads ${DID_P^E}$ from \texttt{clientID} and resolves $DDOC_P^E$. $B$ extracts $PK\_AGG_P^E$ from ${DOC_P^E}$ and uses it with its secret key-agreement key $SK\_AGG_B$ to execute an ECDH-ES key-agreement. The result is the same shared key $K_{P/B}^{ES}$ as used by $P$. $B$ uses $K_{P/B}^{ES}$ to decrypt the \texttt{Authentication Data} $enc({DID_P^S})$ of $M_{con}$ and retrieves $P$'s static long-term ${DID_P^S}$. $B$ resolves $DDOC_P^S$ and extracts $PK\_AGG_P^S$. At this point, $B$ authenticated $P$'s ephemeral $DID_P^E$ since only the subject of $DID_P^E$ has the keys to compute $K_{P/B}^{ES}$. $B$ retrieved $P$'s static long-term $DID_P^S$ and $PK\_AGG_P^S$.

\subsection{Broker requests Authorization Credential}
$B$ uses its secret key-agreement key $SK\_AGG_B$, $P$'s authenticated public ephemeral key-agreement key $PK\_AGG_P^E$ and $P$'s not yet authenticated public static long-term key-agreement key $PK\_AGG_P^S$ to perform an ECDH-1PU key-agreement. The result is the shared key $K_{P/B}^{1PU}$. $B$ generates a secure nonce $N_P$, which is supposed to be used by $P$ for next message encryption to prevent replay attacks and support message ordering. Both, $K_{P/B}^{1PU}$ and $N_P$ are stored by $B$. $B$ prepares an \texttt{AUTH} message $M_{auth}$ and encrypts $N_P$ with $K_{P/B}^{1PU}$. The result $enc({N_P})$ is set as payload $PL$ in $M_{auth}$ and sent to $P$. At this point, $B$ sent $M_{auth}$ as Authorization Credential request to $P$. Since the ECDH-1PU shared key $K_{P/B}^{1PU}$ was used, a MitM cannot decrypt the content.

\subsection{Publisher processes Authorization Credential request}
$P$ receives $M_{auth}$. To decrypt $PL$ from $M_{auth}$, $P$ uses its static long-term secret key-agreement key $SK\_AGG_P^S$, its ephemeral secret key-agreement key $SK\_AGG_P^E$ and $B$'s public key-agreement key $PK\_AGG_B$ to perform an ECDH-1PU key-agreement. The result is the same shared key $K_{P/B}^{1PU}$ as used by $B$. $P$ uses $K_{P/B}^{1PU}$ to decrypt $enc({N_P})$ from payload $PL$, retrieves $N_{P}$ and stores it. If decryption succeeds, $P$ authenticated $B$ as $DID_B$ since only the subject of $DID_B$ has the keys to compute $K_{P/B}^{1PU}$.

\subsection{Publisher sends Authentication Credential}
$P$ will only disclose claims relating to $B$ to prevent $B$ from learning with which other broker $B_{other}$ $P$ is authorized to communicate. $P$ appends to $AC$ the only related disclosure $D$ separated and ending with $"\sim"$. The result is $SDJWT$:
\texttt{<JWT>$\sim<$D$>\sim$}.

$P$ prepares an \texttt{AUTH} message $M_{auth}$ and encrypts $SDJWT$ by using $N_P$ received from $B$ and $K_{P/B}^{1PU}$. The encryption result $enc({SDJWT})$ is set as payload $PL$ to $M_{auth}$. $P$ sends $M_{auth}$ to $B$ and from now on increments $N_P$ after every usage of $K_{P/B}^{1PU}$. This prevents a) replay attacks and b) allows proving the affiliation of multiple message fields.

\subsection{Broker processes Authorization Credential}

$B$ receives $M_{auth}$, loads the previously stored $K_{P/B}^{1PU}$ and $N_P$ and decrypts the $PL$ value $enc({SDJWT})$ from $M_{auth}$ with $K_{P/B}^{1PU}$ and $N_P$ in order to retrieve $SDJWT$. If decryption succeeds, $B$ authenticated $P$ as $DID_P^S$ since only the subject of $DID_P^S$ has the required keys to compute $K_{P/B}^{1PU}$. $B$ splits $SDJWT$ into $AC$ and disclosure $D$ and further splits $AC$ into $JWT$ and $JWT$'s signature $SIG$.

\subsection{Broker processes authorizations}
$B$ verifies that the subject in $JWT$'s \texttt{sub} attribute matches $DID_P^S$ for ensuring that the authorization claims in the $JWT$ are about $P$. $B$ further verifies that the issuers`s $DID_{ISS}$ in the \texttt{iss} attribute is a member of $TIL$. Only then $DID_{ISS}$'s  DID document $DDOC_{ISS}$ is resolved and the public signature verification key $PK\_SIG_{ISS}$ from $DDOC_{ISS}$ is used for the verification of $SIG$. Only if $SIG$ is valid, $B$ uses the credential id from the $JWT$'s \texttt{jti} attribute and checks revocation registry {RR}, whether $JWT$ has been revoked. If $JWT$ is not revoked, $B$ is guaranteed that a) $AC$ is a valid credential issued by a trusted party included in $TIL$, b) $AC$ is about $P$, and c) $AC$ was presented by $P$ since only $P$ could have been able to compute and use $K_{P/B}^{1PU}$. $B$ uses all received disclosures for revealing a) publish authorizations to topics ($\vv{A}_{Pub}$) and b) subscribe authorizations to topics ($\vv{A}_{Sub}$). Since $P$ sent only one disclosure for publishing to $T$, $\vv{A}_{Sub}$ is empty. $B$ stores $\vv{A}_{Pub}$ and $\vv{A}_{Sub}$ and confirms to $P$ the successful connection with a \texttt{CONNACK} message. 

At this point, $P$ is fully authenticated and authorized towards $B$, and $B$ is fully authenticated towards $P$. $P$ and $B$ use $K_{P/B}^{1PU}$ for privacy preserving two-way authenticated encryption of message field content. Authorization data that is not required for authorization by $B$ was not revealed.

\subsection{Subscriber sends connection request}
In order to connect $S$ to $B$, the same steps as described in $A.$ to $H.$ have to be applied for $S$. Additionally, $S$ prepares a \texttt{SUBSCRIBE} message $M_{Sub}$ for subscribing to $T$ at $B$. $S$ encrypts $T$ with its $K_{S/B}^{1PU}$ which results in $enc({T})$. $enc({T})$ is set as payload $PL$ to $M_{Sub}$ which is then send to $B$. $B$ receives $M_{Sub}$, decrypts $enc({T})$ from $PL$ with $K_{S/B}^{1PU}$, and verifies that $T$ is member of the stored subscribe authorizations $\vv{A}_{Sub}$. Only if $T \in \vv{A}_{Sub}$, the subscription of $S$ is acknowledged by $B$ with \texttt{SUBACK}. $S$ is now authorized to receive messages published to $T$ at $B$.

\subsection{Publisher publishes message with encrypted content}
$P$ prepares a \texttt{PUBLISH} message $M_{Pub}$, encrypts the message payload $PL$ and topic $T$ with $K_{P/B}^{1PU}$ and sets it as $\texttt{Payload}$ and \texttt{Topic} of $M_{Pub}$. The nonce used ($N_P$) is incremented after each encryption. $P$ sends $M_{Pub}$ to $B$.

$B$ receives $M_{Pub}$ and decrypts $PL$ and $T$ with $P$'s $K_{P/B}^{1PU}$. $B$ checks that the nonces used for encryption of $PL$ and $TL$ are incremented versions of $N_P$ in order to detect replay attacks or substitution of $PL$ or $T$ by a MitM. $B$ verifies that $T \in \vv{A}_{Pub}$. $B$ encrypts $PL$ and $T$ for $S$ again, but this time with $S$'s $K_{S/B}^{1PU}$ and forwards the message to $S$. $B$ acknowledges to $P$ with \texttt{PUBACK} message. $S$ decrypts the \texttt{PUBLISH} message with $K_{S/B}^{1PU}$ and processes $T$ and $PL$.

\section{Adding and removing system participants}
\label{sec_adding_removing_system_participants}
The following section details how to remove or how to add a new $P$, $S$, $PO$ or $SO$ to a network using DAXiot.

For adding a $PO_{new}$, its $DID_{new}$ needs to be added to $TIL$. 
For adding a $P_{new}$, $PO$ needs to be member of $TIL$ and needs to issue an $AC$ to $P_{new}$.
For removing a $PO$, its $DID_{PO}$ needs to be removed from $TIL$.
For removing $P$, $PO$ marks $P$'s $AC$ as \texttt{REVOKED} in $RR$. 
The same approaches as for adding or removing a $P$ or $PO$ can be applied to $S$ and $SO$. The effort for chaging system participants is low.

\section{Implementation and Evaluation}
\label{sec_implementation_and_evaluation}
We implemented DAXiot in HiveMQ MQTT Client 1.3.0 and HiveMQ CE Broker 2023.3 and used libsodium 1.0.18 for cryptographic operations. We evaluated DAXiot regarding security, privacy, and manageability and executed performance benchmarks. Broker and subscriber were installed on a notebook with Intel Core i5 with 3.8 GHz and 16 GB RAM. The Publisher was installed on a Raspberry Pi 3 Model B.

\textbf{Security Evaluation:}
We achieved confidentiality through authenticated data encryption. Device owners only control own devices. Mutual authorization is not implemented but possible.

\begin{table}[htbp]
\caption{Performance Evaluation and Comparison}
\resizebox{0.5\textwidth}{!}{%
\begin{tabular}{|l|c|c|}
\hline
\textbf{Scenario} & \multicolumn{1}{l|}{\textbf{Connecting (ms)}} & \multicolumn{1}{l|}{\textbf{Message Publishing (ms)}} \\ \hline
Without TLS      & 58.7                                                   & 5.2                                               \\ \hline
Server Side TLS  & 84.4                                                   & 5.6                                              \\ \hline
Mutual TLS       & 91.2                                                    & 5.7                                               \\ \hline
DAXiot           & 115.8                                                    & 6.3                                               \\ \hline
\end{tabular}%
}
\label{tab_performance}
\end{table}

\textbf{Privacy Evaluation:}
Devices are not publicly registered. Device to broker communication is not visible to the public. Authorization data is selectivly disclosed. The broker endpoint must be public due to the nature of our scenario in section \ref{sec_systm_design_and_message_flow}.

\textbf{Manageability Evaluation:}
Our solution is decentralized and is applicable to protocols supporting challenge-response based authentication. Changing system participants comes with low effort and devices do not need to be updated. We do not require additional infrastructure like a blockchain.

\textbf{Performance Evaluation:}
We measured DAXiot's performance and compared it to scenarios without and with (mutual) TLS. We established 1000 connections and published 10000 messages and calculated the average execution time required by the publisher (Table \ref{tab_performance}). With DAXiot, publishing was about 10\% slower than using mutual TLS, with connection establishment taking roughly 21\% longer. However, one must take into account that connecting with DAXiot includes authorization. Though we consider the decrease in performance negligible when compared to the security and privacy gains offered by our solution.

\section{Conclusion and Future Work}
\label{sec_conclusion_and_future_work}
With DAXiot we propose a decentralized and privacy preserving authentication and authorization scheme for dynamic IoT networks. We applied DAXiot to an MQTT 5.0 scenario in-band, which differentiates DAXiot from other solutions in which authentication and authorization with DIDs and VCs are commonly performed out-of-band and require additional infrastructure components (e.g. blockchain). We enable authenticated encryption solely based on DIDs and disclose only necessary authorization data. Eavesdropper are not able to read encrypted content. Participants in IoT networks can be switched with minimal effort, without requiring device updates. Benchmarks proved adequate performance for an application of DAXiot in cooperative and dynamic IoT networks.

For future work we plan to a) investigate scenarios in which network operators instead of device owners give permissions to devices and b) extend our solution with mutual authorization. Further we intend to investigate the applicability of public keys of DIDs for key agreements to enable TLS instead of encrypting individual message fields. This would facilitate a more seamless integration of DAXiot into common protocols supporting TLS.


\begin{thebibliography}{00}
\bibitem{bib_schiller_landscape_iot_security}
E. Schiller, A. Aidoo, J. Fuhrer, J. Stahl, M. Ziörjen, and B. Stiller, “Landscape of IoT security,” Computer Science Review, vol. 44, May 2022

\bibitem{bib_mqtt}
OASIS, "MQTT Version 5.0," 2019, https://docs.oasis-open.org/mqtt/mqtt/v5.0/mqtt-v5.0.html

\bibitem{bib_maigne_centralized_distributed_between_IoT_Security}
S. Dramé-Maigné, M. Laurent, L. Castillo, and H. Ganem, “Centralized, Distributed, and Everything in between: Reviewing Access Control Solutions for the IoT,” ACM Comput. Surv., vol. 54, no. 7, pp. 1–34, Sep. 2022

\bibitem{bib_industry40_state_of_art}
L. D. Xu, E. L. Xu, and L. Li, “Industry 4.0: state of the art and future trends,” International Journal of Production Research, vol. 56, no. 8, pp. 2941–2962, Apr. 2018

\bibitem{bib_ravidas_AC_in_IoT_survey}
S. Ravidas, A. Lekidis, F. Paci, and N. Zannone, “Access control in Internet-of-Things: A survey,” Journal of Network and Computer Applications, vol. 144, Jul. 2019

\bibitem{bib_abubakar_blockchain_based_MQTT} M. Abdelrazig Abubakar, Z. Jaroucheh, A. Al-Dubai, and X. Liu, “Blockchain-based identity and authentication scheme for MQTT protocol,” in 2021 The 3rd International Conference on Blockchain Technology, Shanghai China: ACM, pp. 73--81, Mar. 2021.

\bibitem{bib_dixit_decentralized_it_framework} A. Dixit, M. Smith-Creasey, and M. Rajarajan, “A Decentralized IIoT Identity Framework based on Self-Sovereign Identity using Blockchain,” in 2022 IEEE 47th Conference on Local Computer Networks (LCN), Edmonton, AB, Canada: IEEE, pp. 335--338, Sep. 2022. 

\bibitem{bib_belchior_ssibac} R. Belchior, B. Putz, G. Pernul, M. Correia, A. Vasconcelos, and S. Guerreiro, “SSIBAC: Self-Sovereign Identity Based Access Control,” in 2020 IEEE 19th International Conference on Trust, Security and Privacy in Computing and Communications (TrustCom), Guangzhou, China: IEEE, pp. 1935--1943, Dec. 2020.

\bibitem{bib_fotiou_capabality_based_access} N. Fotiou, V. A. Siris, G. C. Polyzos, Y. Kortesniemi, and D. Lagutin, “Capabilities-based access control for IoT devices using Verifiable Credentials,” in 2022 IEEE Security and Privacy Workshops (SPW), pp. 222--228, May 2022.

\bibitem{bib_w3c_did_core}
M. Sporny, D. Longley, M. Sabadello, D. Reed, O. Steele, and C. Allen, “Decentralized identifiers (dids) v1.0,” W3C, Recommendation, Jul. 2022, https://www.w3.org/TR/did-core/.

\bibitem{bib_w3c_did_specification_registries}
O. Steele, and M. Sporny, “DID Specification Registries,”, W3C, May 2023, https://www.w3.org/TR/did-spec-registries/ 

\bibitem{bib_did_ethr_method}
“ethr DID Resolver,” Decentralized Identity Foundation, Nov. 2022, https://github.com/decentralized-identity/ethr-did-resolver

\bibitem{bib_did_web_method}
M. Prorock, O. Steele, and O. Terbu, “did:web Method Specification,” https://w3c-ccg.github.io/did-method-web/ 

\bibitem{bib_did_key_method}
D. Longley, D. Zagidulin, and M. Sporny. "The did:key method,", Sep. 2022, https://w3c-ccg.github.io/did-method-key/.

\bibitem{bib_w3c_vc_data_model_11}
M. Sporny, D. Longley, and D. Chadwick, “Verifiable Credentials Data Model v1.1,”, W3C, Recommendation, March 2022, https://www.w3.org/TR/vc-data-model/ 

\bibitem{bib_vc_flavors_Article}
K. Young, “Verifiable Credentials Flavors Explained,”, Feb. 2022, https://www.lfph.io/wp-content/uploads/2021/02/Verifiable-Credentials-Flavors-Explained.pdf

\bibitem{bib_anoncreds_specification}
S. Curran, A. Philipp, H. Yildiz, S. Curren, and V. M. Jurado , “AnonCreds Specification,”, 2022, https://hyperledger.github.io/anoncreds-spec/

\bibitem{bib_sd_jwt}
D. Fett, K. Yasuda, and B. Campbell, “Selective Disclosure for JWTs (SD-JWT),” Internet Engineering Task Force, Jun. 2023, https://datatracker.ietf.org/doc/draft-ietf-oauth-selective-disclosure-jwt

\bibitem{bib_elliptic_curve_cryptography}
D. Hankerson, A. Menezes, "Elliptic curve cryptography," in Encyclopedia of Cryptography, Security and Privacy, pp. 1-2. Berlin, Heidelberg: Springer Berlin Heidelberg, 2021

\bibitem{bib_Diffie_Hellman_Key_Exchange} N. Li, “Research on Diffie-Hellman key exchange protocol,” in 2010 2nd International Conference on Computer Engineering and Technology, Apr. 2010. 

\bibitem{bib_state_of_elliptic_curve_cryptography}
N. Koblitz, A. Menezes, and S. Vanstone, “The State of Elliptic Curve Cryptography,” Designs, Codes and Cryptography, vol. 19, no. 2, pp. 173–193, Mar. 2000.

\bibitem{bib_curve25519} 
D. J. Bernstein, “Curve25519: New Diffie-Hellman Speed Records,” in Public Key Cryptography - PKC 2006, M. Yung, Y. Dodis, A. Kiayias, and T. Malkin, Eds., in Lecture Notes in Computer Science, vol. 3958. Berlin, Heidelberg: Springer Berlin Heidelberg,  pp. 207–228, 2006.

\bibitem{bib_conversion_ed25519_x25519}
E. Thormarker, “On using the same key pair for Ed25519 and an X25519 based KEM,”, 2021,  https://eprint.iacr.org/2021/509

\bibitem{bib_ecdh1pu}
N. Madden, “Public Key Authenticated Encryption for JOSE: ECDH-1PU,” Internet Engineering Task Force, May 2021, https://datatracker.ietf.org/doc/draft-madden-jose-ecdh-1pu-04

\bibitem{bib_multibase}
J. Benet and M. Sporny, “The Multibase Data Format,” Feb. 19, 2023, https://w3c-ccg.github.io/multibase/

% \bibitem{bib_raw_public_keys}
% P. Wouters, H. Tschofenig, J. I. Gilmore, S. Weiler, and T. Kivinen, “Using Raw Public Keys in Transport Layer Security (TLS) and Datagram Transport Layer Security (DTLS),” Internet Engineering Task Force, Request for Comments RFC 7250, Jun. 2014.

\end{thebibliography}
\end{document}